\documentclass[conf]{new-aiaa}
\usepackage[utf8]{inputenc}

\usepackage{graphicx}
\usepackage{float}
\usepackage{caption}
\usepackage{subcaption}
\usepackage{amsmath}
\usepackage[version=4]{mhchem}
\usepackage{siunitx}
\usepackage{longtable,tabularx}
\usepackage{multirow}
\usepackage{stackengine}
\newcommand\xrowht[2][0]{\addstackgap[.5\dimexpr#2\relax]{\vphantom{#1}}}
\title{Insights to the Forensic Investigation of a Custom Built UAV}

\author{Tanay Kumar \footnote{Research Associate, Department of Aerospace Engineering; ktanay@iitk.ac.in}, Mangal Kothari\footnote{Professor, Department of Aerospace Engineering; mangal@iitk.ac.in, Member AIAA.}}
\affil{Indian Institute of Technology Kanpur, Kanpur, UP, India, 208016}

\begin{document}

\maketitle

\begin{abstract}
Unmanned Aerial Vehicles (UAVs) have revolutionized numerous application fields since their introduction. Alongside this, advancements in mechanics and electronics have simplified and decreased the cost of system design. As a consequence, UAVs have dominated the global market across all sectors from recreational products to military applications. However, the widespread use of UAVs has also contributed to an increase in criminal activity involving them. UAVs have become a common means of delivering narcotics, arms, and ammunition, collecting sensitive personal information, and snooping in restricted areas, among other activities. This could constitute a threat to national security. Despite its growing significance, UAV forensics is still a relatively unexplored field of study. In this paper, we present new insights on UAV forensic analysis in terms of preliminary analysis, accessing the digital containers of the UAV, and retrieving vital data. With the aid of a simulated scenario, a comprehensive methodology for the analysis of retrieved data to establish conclusions on the flight path, source of origin, recover flight data, and acquire media file content is presented.
\end{abstract}



\section{Introduction}

\subsection{Drones in the modern world}
Unmanned Aerial Vehicles (UAVs) have grown in popularity in recent years and have become an integral part of numerous industries, such as agriculture, media, recreation, disaster management, delivery, the military, etc\cite{shakhatreh2019unmanned, mondal2020autonomous, muhammad2020vocal}. In parallel with this, advancements in mechanics and electronics have simplified and decreased the cost of system design. UAVs have revolutionized the collection of data and execution of duties in the air.  They have the unique ability to deploy from anywhere and to transport a variety of payloads.  Due to these benefits, they have become a popular area of research and consequently, tech giants such as Amazon, Google, Intel, and Boeing are investing heavily in research and development. 

Nonetheless, the proliferation of UAVs has posed new difficulties for law enforcement and forensic investigators. UAVs have been used to commit offenses such as contraband smuggling, illegal surveillance, and commercial flight interference\cite{sathyamoorthy2015review}. In the recent past, drones have also been captured inadvertently violating no-fly zones, and there is a growing fear that terrorists could use them to cause panic or other damage. Therefore, an effective and expedient method of investigating UAV-related incidents is required. The potential misuse of UAVs in criminal and terror-related activities has intrigued the interest of forensic analysts. Due to their vertical takeoff-and-landing (VTOL) and hover capabilities, multirotor UAVs have become popular for malicious purposes. Small and nimble, they are ideal for conducting surveillance and reconnaissance in secret. They can also contain payloads such as cameras or small packages, allowing them to transport contraband or illegal goods across borders or into restricted areas. Furthermore, these UAVs are widely available, relatively affordable, and simple to operate, making them accessible to individuals with malicious intent. It is important to note, however, that any form of UAV can be used for illegal activities if operated improperly, and the vast majority of UAV operators use them for legal and beneficial purposes.

\subsection{Motivation and aim}
Due to the threats posed by the malicious use of UAVs, UAV forensic programs are essential for advancing our understanding of drone technology. In addition, UAV forensics can be utilized for crash analysis, which is prevalent in the research and development industry. There is only little work in this area; therefore, UAV forensics is an underexplored discipline. In order to address the lack of a standardized drone-based collection and analysis model in the field of forensics, \cite{alotaibi2022comprehensive} highlights the extant forensic models proposed for drone forensics and proposes a novel comprehensive collection and analysis forensic model applicable to drone forensics. A generic drone forensic model that improves the digital investigation procedure for general UAV forensics applicable to a broader spectrum of UAVs has been proposed in \cite{jain2017drone}. In addition, the paper describes the forensic analysis model of the various components, including a camera and Wi-Fi. In \cite{bouafif2018drone}, the authors provide new insights into drone forensics in terms of accessing the digital contents of an intercepted drone and retrieving all the data that can assist digital forensic investigators in establishing ownership, recovering flight data, and acquiring media file content. The results of the examination of the exposed FTP channel and data extraction from the memory chip of the DJI Mavic Air 2 Drone are discussed in \cite{lan2022drone}. In the case study \cite{kao2019drone}, the collection, examination, correction, and analysis of vital data from a DJI Spark drone's recorded flight data are described. The investigation of multiple drones produced by industry-leading manufacturers such as DJI, Bebop, and Syma is done in \cite{al2021drone}. Extracted are forensically pertinent data such as location, captured images and videos, UAV's flight paths, and ownership information. In addition, the paper establishes results to aid law informants in this domain. A study on the identification of what data can be extracted from UAV devices is examined in \cite{atkinson2021drone}. Further, the utility of the extracted data and whether consumers can obfuscate the data in an effort to evade detection using data collected from UAV flight testing have also been discussed.

Although UAV forensics has attracted the attention of researchers and forensic analysts in recent years, the vast majority of studies are conducted using commercially available drones such as DJI or Parrot.  Using a serial number, commercial UAVs are registered with the company and frequently with the government as well. Such UAVs may also include a capability for storing flight history in the cloud. These features facilitate the tracking of the UAV's proprietor and the retrieval of any other data stored on the UAV. Moreover, such UAVs are costly because of the advanced technology, quality components, and manufacturing costs. Due to these factors and the abundance of data available online for the design and manufacture of UAVs, drone operators have begun to employ custom-made UAVs for malicious purposes. The use of custom technologies and components enables the operator to customize the UAV, particularly its software, in a way that may not be possible with commercial UAVs. This makes identification and tracking of the UAV difficult. Consequently, it is necessary to examine the scope of UAV forensics in such systems. This research paper provides a comprehensive methodology of UAV forensics for collecting and analyzing digital evidence from UAVs with custom-built components. The study examines a custom-made UAV, its components, and the methodologies for data extraction. In addition, the paper discusses the processing and analysis of extracted data in order to draw conclusions. The findings contribute to the development of effective investigative techniques for UAV-related incidents and provide valuable insights into the field of UAV forensics. 

The remainder of the is structured as follows: Section \ref{UAV} provides an overview of UAVs, their classification, design and development, and payload details. Section \ref{Problem} establishes the groundwork for the problem statement. This section discusses the current threats posed by UAVs, the significance of UAV forensics, the evidence and data available to combat these threats, and the difficulties associated with UAV forensics. Using a simulated scenario, Section \ref{Problem} presents a methodological approach to the UAV forensics problem. The latter portions of the section discuss the data retrieval processes, the analysis, and the obtained results. In Section \ref{Conclusion}, concluding remarks are presented.

\section{Unmanned Aerial Vehicles}
\label{UAV}

\subsection{Classification of UAVs}
It is imperative to classify UAVs due to multiple reasons. Classification helps to standardise the various types of available UAVs, making it easier to comprehend their capabilities, specifications, and applications.
It enables operators to select the most appropriate UAV for a particular application based on size, range, payload capacity, and endurance, ensuring that the UAV is optimized for the task at hand. Classification also aids regulators in the development and implementation of safety and operational standards for various UAV categories. UAVs can be categorized based on the parameters listed below.

\begin{enumerate}
\item \textbf{Size:} UAVs can be classified as micro, small, medium, and large based on their size. Micro UAVs typically weigh just a few grams, whereas small UAVs can weigh up to 25 kg. Medium UAVs can weigh up to 250 kg, and large UAVs can weigh over 500 kg. Each size classification of UAVs has its unique features, capabilities, and applications. Micro and small UAVs are ideal for indoor use and aerial photography, while medium and large UAVs are more suitable for long-range surveillance and military operations.

\item \textbf{Range:} UAVs can also be classified based on their range. Short-range UAVs can fly up to 50 km from their launch point, while medium-range UAVs can travel up to 150 km. Long-range UAVs can travel over 150 km from their launch point, and some have an endurance of over 36 hours. The range of UAVs depends on factors like size, propulsion systems, and fuel capacity. 

\item \textbf{Application:} UAVs have numerous uses, including military, civilian, and commercial applications. The main tasks of military UAVs are reconnaissance, surveillance, and combat missions. UAVs are also utilized for search and rescue operations, disaster management, and environmental monitoring in the civilian sector. Commercial UAVs are used in industries such as agriculture, construction, and surveying. UAVs utilized for aerial photography and videography are also gaining popularity. Each UAV application demands unique characteristics, including payload capacity, range, and endurance. 

\item \textbf{Configuration:} There are two primary configurations of UAVs: fixed-wing and rotary-wing. Fixed-wing UAVs are built similarly to conventional aircraft. They are typically employed for long-range missions and are can fly at high altitudes and velocities for extended periods of time. Alternatively, rotary-wing UAVs are capable of vertical takeoff and landing. They are typically employed for shorter-range missions and have hover capability. These two configuration types also include subcategories depending on the number of rotors, the design of the wing, etc.  In addition to this, hybrid UAVs, which incorporate characteristics of fixed-wing and rotary-wing UAVs, are gaining popularity due to their adaptability and ability to take off vertically and fly forward like fixed-wing UAVs. Further, flapping-wing UAVs are also gaining popularity. However, they have not matured enough for commercial use. 
\end{enumerate}

\subsection{Components of UAVs}
The UAV components depend solely on their configuration, size, and mission. However, as stated previously, quadrotor UAVs are the most common, so we will focus primarily on quadrotor UAV components. Each element on any UAV can be identified as one of the following despite the fact that not every UAV will contain all of the enumerated components.

\begin{enumerate}
    \item \textbf{Frame and chassis:} The frame is typically made of lightweight materials such as carbon fiber and acts as the core fuselage to house all the other components.
    
    \item \textbf{Motors and propellers:} A quadrotor has a set of four motors and propellers to provide the lift and propulsion for the UAV. The diagonally opposite rotors always spin in the same direction to control the pitch, roll, and yaw motion. Long range and/or high endurance multi-rotor can have more than four rotors. Among all multi-rotors, quadcopters, hexacopter, and octarotor are very popular. 

    \item \textbf{Electronic Speed Controllers (ESCs):} The ESCs control the speed of the motors and the propellers. ESCs generate the desired thrust by driving motors to the specified speed using a specialized combination of hardware and firmware. 

    \item \textbf{Flight controller:}  The flight controller is the brain of the UAV and is responsible for the control and stability of the UAV. It uses sensor data (gyroscopes, accelerometers, and magnetometers) combined with GPS for autonomous navigation. In addition to this, the vehicle can be controlled remotely by radio control. 

    \item \textbf{GPS receiver:} This component is required for autonomous flight in order to manage UAV position and return to home functionality. It also helps to autonomously pilot the UAV beyond the visual line of sight (BVLOS).

    \item \textbf{Radio receiver:} Used to receive control input signals sent by the pilot using a ground based transmitter

    \item \textbf{Transmitter:} Transmits manual input from the operator on the ground to the UAV.

    \item \textbf{Battery:} The battery provides power to the motors, the flight controller, and other electronic components of the UAV.
    
    \item \textbf{Telemetry:} The telemetry module permits bidirectional communication between the UAV and the ground station. All of the UAV's real-time data can be transmitted to the ground station for monitoring, allowing the operator to act accordingly and send commands back to the UAV.
    
    \item \textbf{Ground Control Software (GCS):} The GCS allows the operator to remotely control and monitor the UAV, including its position, attitude, and speed. It also provides information about the UAV's vital status, such as battery and sensor data, enabling efficient mission planning and execution. This software is used to control pre-determined navigation and effectively plan flight missions.  
\end{enumerate}

\begin{figure}[ht]
\centering
 \includegraphics[scale = 0.4]{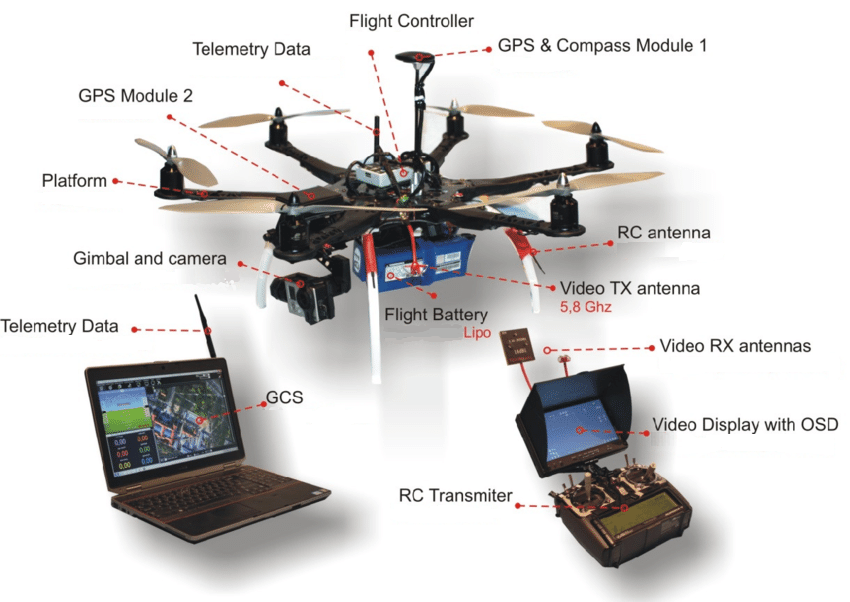}
 \caption{Components of a typical multirotor UAV\cite{burdziakowski2017low}}
 \label{fig:Component}
\end{figure}

\subsection{UAV Payload}
UAVs are versatile platforms that can be equipped with various types of payloads depending on the intended application. Some common types of payloads are listed below.

\begin{enumerate}
    \item \textbf{Imaging and sensing payload:} UAVs can be equipped with various types of cameras and sensors, such as RGB cameras, thermal cameras, LiDAR sensors, and multispectral sensors. These payloads are used for aerial photography, surveying, inspection, mapping, geo-tagging, and remote sensing.

    \item \textbf{Cargo:} In recent years, the use of UAVs to provide timely and efficient deliveries has become an area of increasing investment, with Amazon's Prime Air service garnering the most attention. While the primary emphasis is on commercial and retail deliveries, this technology could also benefit other industries such as healthcare.

    \item \textbf{Scientific instruments:} UAVs can be used to carry scientific equipment like atmospheric and biological sensors for research and monitoring applications.

    \item \textbf{Weapons:} UAVs may be equipped with combative weapons for transportation or to conduct attacks at the desired target location. 

    \item \textbf{Electronics and communication payload:} Communications payloads are not yet prevalent, but they have attracted considerable interest in recent years. UAVs can be used to carry jammers, signal spoofers, and other signal intelligence systems.
\end{enumerate}

\section{Problem Overview}
\label{Problem}

\subsection{Evidence Sources and Data Retrieved}
Various forms of data and evidence can be retrieved from the UAV. Proper analysis of the evidence can help track its source. Some of the sources of evidence in drone forensics include:

\begin{enumerate}
    \item \textbf{Video and image data:} Reconnaissance and spying drones always carry cameras. In most instances, captured images and videos are stored onboard and can serve as a primary data source. This data can be analyzed to provide evidence of the flight path of the drone, its location, and any nearby objects or persons.

    \item \textbf{Other payload content and storage devices:} Secondary storage devices may be present on the UAV to record huge quantities of sensor data. This data is frequently geo-tagged and contains other crucial data, when analyzed, can yield vital information.

    \item \textbf{Cloud-based data platforms:} This type of storage is uncommon at present, but it is gaining popularity as 5G technology advances. The user could use cloud data to reduce the need for local storage. 

    \item \textbf{Flight logs:} Majority of the drones are equipped with flight loggers that record all information about the drone's flight. These flight records can be used to reconstruct the drone's flight path and identify its source location. 

    \item \textbf{GPS data:} The GPS receivers used in drones often have a memory of their own to record the real-time GPS data. The logging feature allows position fixes and arbitrary byte strings from the host to be logged in flash memory attached to the receiver. This is incredibly useful when data logging is manually disabled on the flight controller of a custom drone. The data can be utilized to analyze the flight path of the vehicle.  

    \item \textbf{Network packet data:} The UAV wirelessly communicates with the ground station using a communication protocol (Mavlink is popular). The data packets can be intercepted and decrypted to obtain information about the drone's flight path and other vitals. Reading the data packets can, however, be a challenging task as they are encrypted with the AES encryption key.

    \item \textbf{Physical evidence:} Components of drones, such as propellers, batteries, and other elements, can be analyzed to determine the type of drone, the manufacturer, and other useful information. Batteries and propellers can provide an approximation of the drone's range, which when combined with other evidence, can help identify its origin. Components and payload utilized may have a serial number that can be tracked.  Typically, commercially available drones such as DJI and Parrot sync previous flight logs and other data to cloud-based platforms that can be accessed using the drone's serial number. Additionally, the drone and its associated equipment may contain moist evidence such as fingerprints, DNA, etc.

    \item \textbf{Witness testimonies:} Witnesses who saw or heard the drone flying can provide valuable information about the drone's flight path and the location of the drone.

\end{enumerate}

\subsection{Current Threats and Offences}
\begin{figure}[ht]
\centering
\begin{minipage}{.45\textwidth}
 		\centering
            \includegraphics[scale = 0.13]{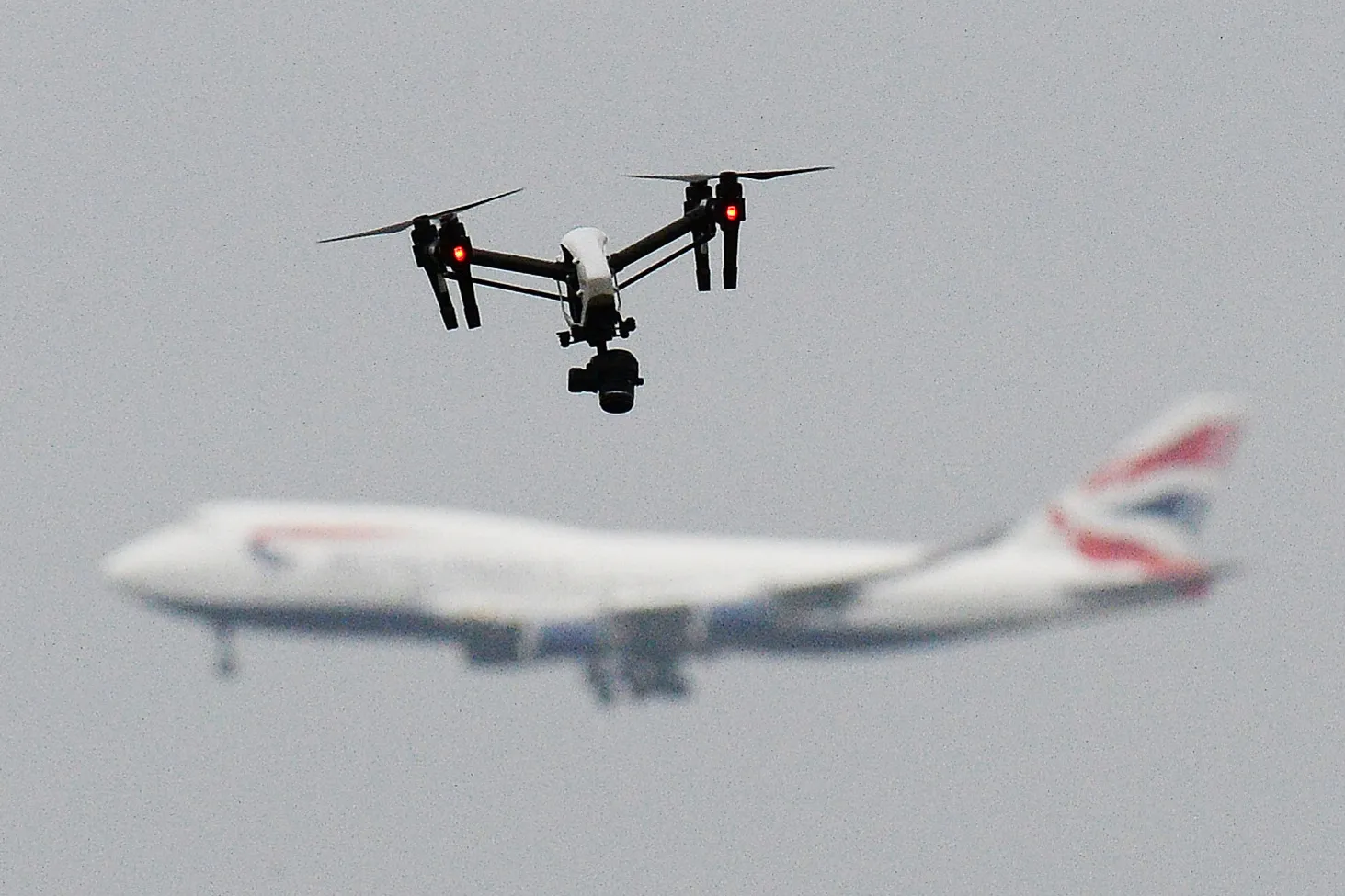}
            \caption[]{UAV caught spying around the \\London Gatwick airport in 2018 \footnotemark[1]}
            \label{fig:Gatwick} 
            
\end{minipage}
\begin{minipage}{.45\textwidth}
             \centering
             \includegraphics[width=0.9\textwidth]{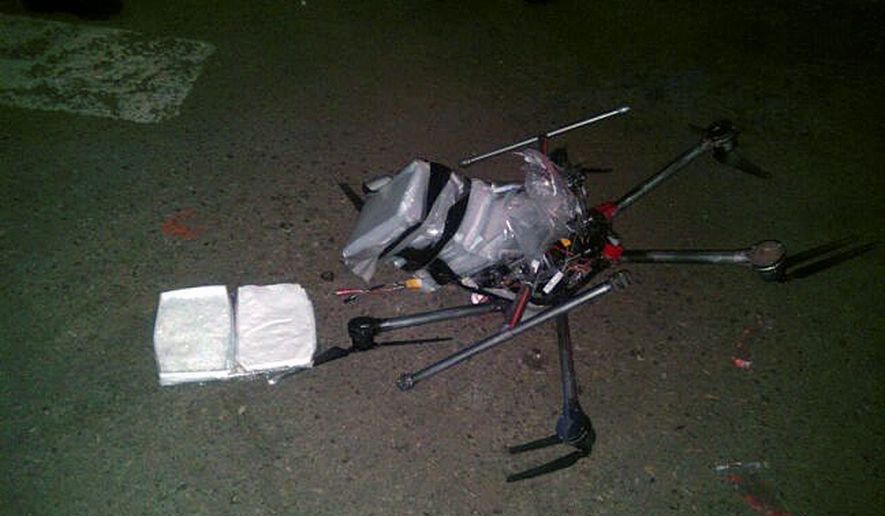}
 	      \caption[]{Mexican drug cartels using drones to \\smuggle narcotics into the U.S. in 2015 \footnotemark[2]}
 	      \label{fig:Mexico}
\end{minipage}
\end{figure}

\begin{figure}[ht]
\centering
 \includegraphics[width=0.4\textwidth]{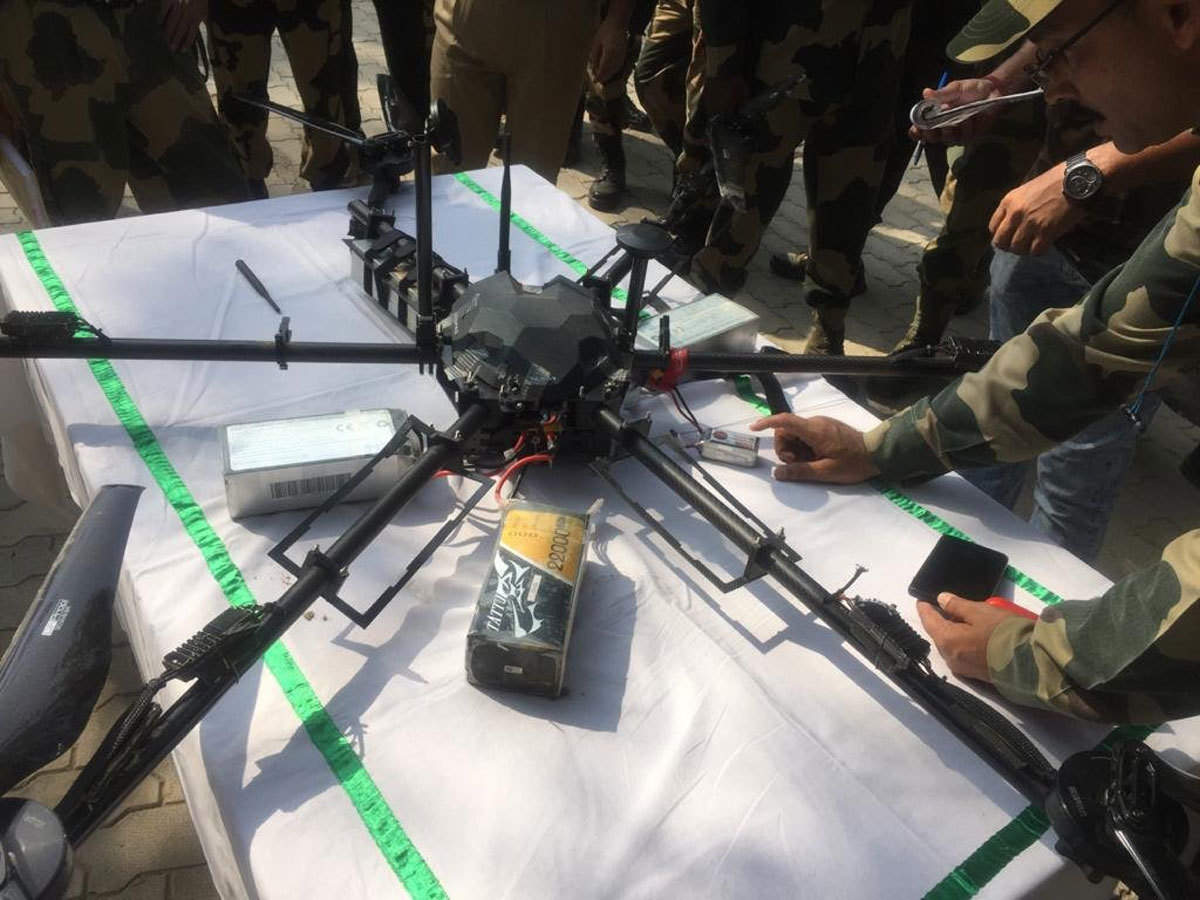}
 \caption[]{Pakistani drone carrying arms shot down by the Indian army in 2020\footnotemark[3]}
 \label{fig:JK}
\end{figure}

With the advancement of technology, UAVs have become a widely available commodity including in the form of toys for children. Drones are part of everyday life, from recreation to the media and entertainment. However, as with any other form of technology, if not handled properly, UAVs can be used for malicious purposes, posing a threat to the general public. The increased availability of UAVs to the general public has also led to their use in illegal activities. Multiple instances of privacy invasion by spy drones have been reported. They are used to collect sensitive personal information such as credit card information and private communications. Additionally, UAVs have been observed in restricted or prohibited areas, such as airports and government buildings, where they pose a safety risk. As per a news article dated December 21, 2018, numerous UAVs were spotted near the runway of Gatwick Airport (Fig.~\ref{fig:Gatwick}), because of which all the aircraft activities had to be shut down \cite{gatwick}. This was not only a security threat but also resulted in the disruption of flight operations and a loss in revenue. Drones have become a common method for smuggling narcotics, arms, and ammunition across international borders, as well as aiding terrorist operations. In January 2015, a drug-toting drone crashed into a supermarket parking lot in Mexico, a couple of miles from the U.S. border shown in Fig.~\ref{fig:Mexico}. The article states this was not the first incident of such a kind \cite{Mexico}. Another instance of UAVs carrying explosives was caught by the BSF in Srinagar, India in February 2021 \cite{JK}. The UAV shown in Fig.~\ref{fig:JK} was carrying small yet powerful IEDs and could be attached to vehicles and detonated remotely. UAVs with internet connectivity can be hacked and used for malicious purposes, such as data theft or cyberattacks. Although extremely uncommon, UAVs can be armed to conduct targeted assassinations or infrastructure attacks. Unauthorised or improperly monitored UAV use in public areas can pose security risks such as explosives and chemical attacks. 

\renewcommand{\thefootnote}{\arabic{footnote}}

\footnotetext[1]{Image credits: \copyright Wired.co.uk /John Stillwell/PA Wire/PA Images}
\footnotetext[2]{Image credits: AP Photo/Secretaria de Seguridad Pública Municipal de Tijuana}

The aforementioned events are common occurrences that made headlines. The use of UAVs in illicit activity is limitless due to the escalating technological advances and falling prices of UAVs. This poses the law enforcement community with new challenges. As a result of such threats, drone forensic programmes are crucial for advancing the comprehension of drone technology, lowering the crime rate, and ensuring the safety and security of individuals, organisations, and the government. Not always is a criminal activity associated with UAV forensics. Accidents involving UAVs are extremely prevalent in the research and development industry, and drone forensics can help determine the cause and parties involved. This can help prevent future accidents of a similar nature. There are restrictions on where drones can fly and how they can be used in a number of nations. Using drone forensics, it is possible to determine whether or not these regulations have been violated and to identify the offenders. Despite the benefits a drone forensics programme can provide to law enforcement agencies, this is still a relatively unexplored area \cite{bouafif2018drone}. 

\footnotetext[3]{Image credits: File photo taken from The Times of India}

\subsection{Forensic Challenges}
The reliability of the analysis and, consequently, its conclusions are highly dependent on the quality and quantity of the evidence gathered. Analysis of UAVs for forensic purposes presents numerous challenges. Collecting relevant data from drones, which can fly at high altitudes and are often difficult to track, is one of the greatest obstacles. Drones can collide or be destroyed during flight, making it difficult to retrieve data or any other physical evidence. Typically, drones employ encryption to safeguard data, making it difficult to extract pertinent information. Encryption also makes it difficult to decode the data packets exchanged between the UAV and the ground station using the telemetry module. Forensic investigators must possess the skills and equipment necessary to decrypt data without damaging it. In addition, modern drones are intricate machines with sophisticated software and hardware components, necessitating an in-depth comprehension of technical details by investigators. Drones may capture images or other data that violates an individual's privacy, raising privacy concerns. Lastly, as drone technology continues to evolve, forensic investigators must keep abreast of the most recent advancements in the field.

\section{Analysis and Results}
\label{Analysis}
\begin{figure}[ht]
	\centering
	\includegraphics[scale= 0.8]{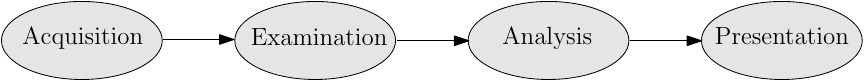}
	\caption{UAV forensic analysis model}
	\label{fig:model}
\end{figure}

In this section, a comprehensive method for the forensic analysis of UAVs is outlined using a simulated scenario. Notably, acquiring and analysing evidence and data from drones can be a complex process requiring specialised skills and equipment. The various components of UAV forensics enable law enforcement agencies to construct an extensive understanding of how drones have been used, the evidence they contain, and even new discoveries regarding drone data. For thorough examination and analysis of the evidence, it is necessary to adhere to the chronological order depicted in Fig.~\ref{fig:model}.

Acquisition is the process of gathering and organising all physical and non-physical data that can be utilized for analysis. During the examination and analysis phase, collected data is observed and evaluated to draw conclusions. Each of these steps can be repeated until conclusive results are reached. The presentation phase necessitates presenting findings in a manner that authorities can comprehend. After completing the analysis phase, the examiner must compile the findings and results into a forensic report.

\subsection{Simulation Scenario}
\begin{figure}[H]
	\centering
	\includegraphics[scale= 0.15]{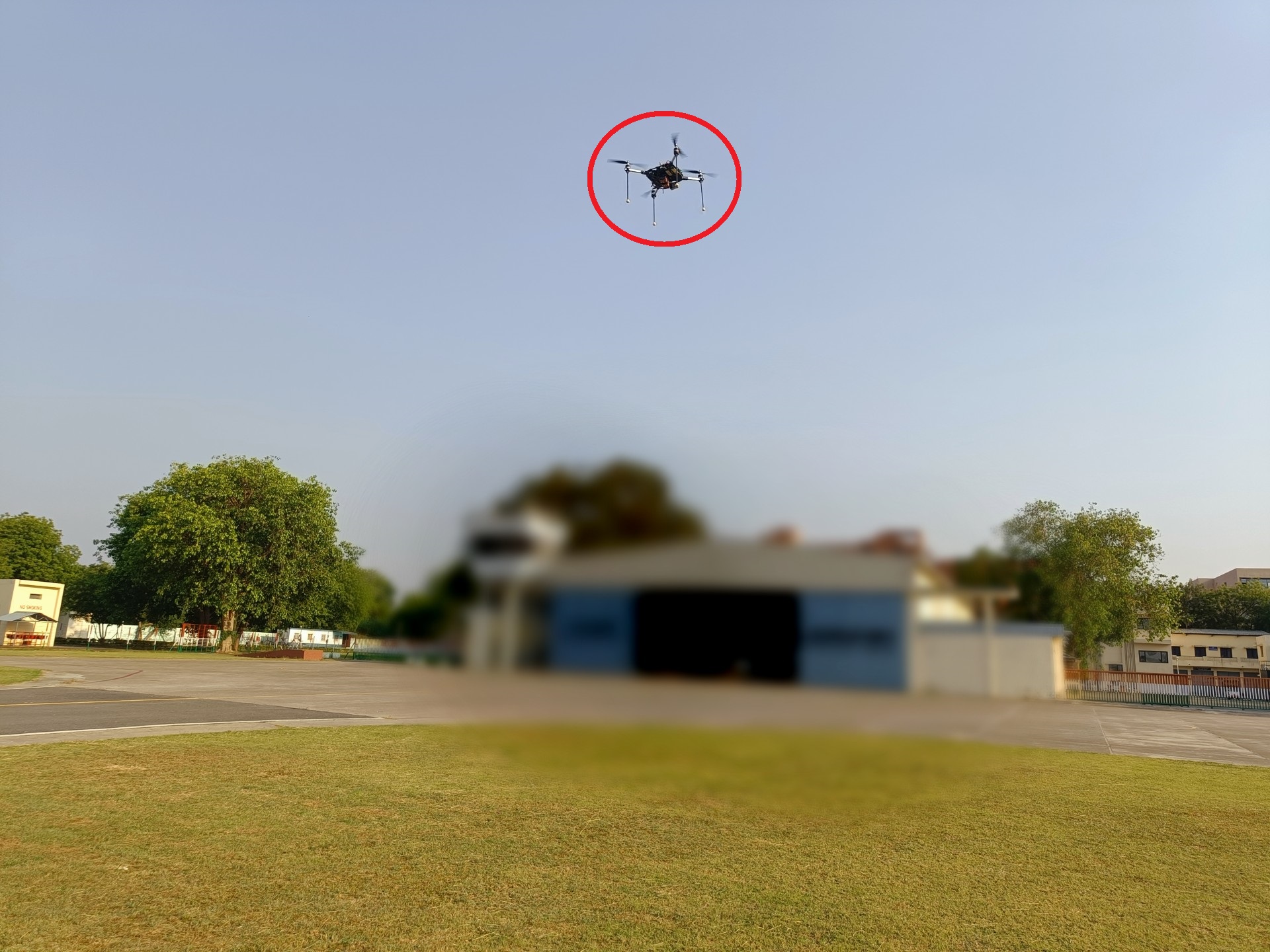}
	\caption[]{UAV intercepted while spying in restricted area \footnotemark[4]}
	\label{fig:Intercept}
\end{figure}

To mimic a situation in which a drone is discovered conducting reconnaissance in a sensitive area, a fictitious scenario is simulated. The drone is immediately identified and apprehended, as depicted in Fig.~\ref{fig:Intercept}. All physical evidence is collected meticulously and sent for forensic analysis.  A custom drone is used for this purpose as it is challenging to retrieve and analyze the data as compared to commercially available drones such as DJI. The flight test is carried out at the Flight Laboratory, IIT Kanpur. What follows is comprehensive forensics of the evidence.

\footnotetext[4]{Image blurred due to security concerns.}

\subsection{Physical Examination}
\label{Physical Examination}
\begin{figure}[H]
\centering
\begin{subfigure}{.49\textwidth}
  \centering
  \includegraphics[scale= 0.3,]{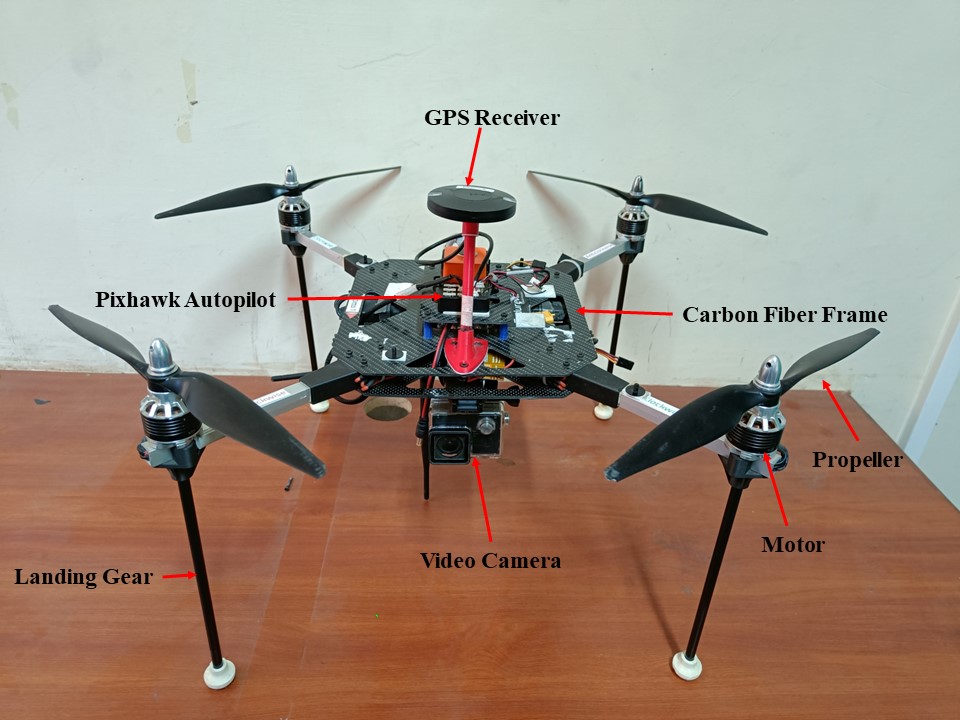}
  \label{fig:Quad1}
\end{subfigure}
\begin{subfigure}{.49\textwidth}
  \centering
  \includegraphics[scale= 0.3,]{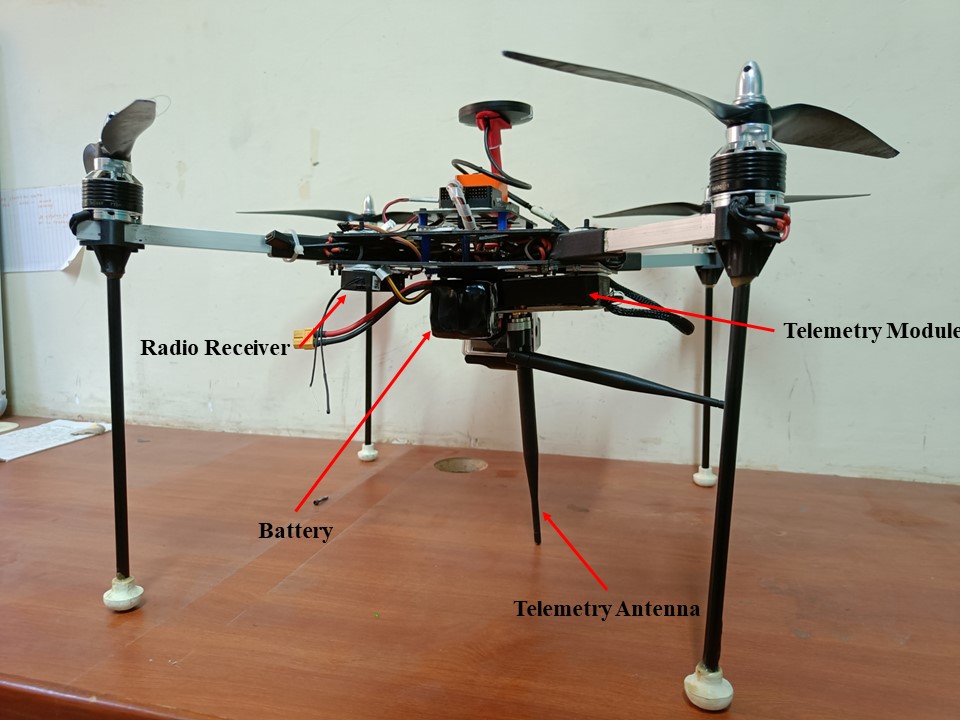}
  \label{fig:Quad2}
\end{subfigure}
\caption{Physical examination of the retrieved UAV}
\label{fig:Quad}
\end{figure}

Physical examination of a drone in UAV forensics typically involves inspecting the drone for any physical damage, collecting data from the drone's internal components and analyzing the collected data to understand the drone's behavior and any potential tampering. The physical evidence is collected and sent to the laboratory for careful examination. The images of the UAV are given in Fig.~\ref{fig:Quad}. Examination of the exterior revealed no form of tampering or sign of wet evidence. The UAV was then carefully dismantled and the components were studied. A list of all the physical components present on the UAV are given in Table~\ref{table:Components}.

\begin{table}[ht]
	\caption{{Physical components retrieved from the UAV}} 
	\centering
	\begin{tabular}{|c|c|}
         \hline
		\textbf{Component} & \textbf{Description}\\ \hline
        Frame & Custom designed carbon fibre frame with `X' quadrotor configuration\\ \hline
        Motors & KDE 2814XF 775KV\\ \hline
        ESC & KDEXF-UAS35\\ \hline
        Propellers &  12$\times$6 hard plastic\\ \hline
        Battery &  LiPo 4s 6500mAh\\ \hline
        Autopilot & Pixhawk cube orange \\ \hline
        GPS Module & Here2 GPS\\ \hline
        Radio Receiver & FrSky X8R capable of long range communication\\ \hline
        Telemetry module & RFD900\\ \hline
        Camera & SJCAM SJ4000 HD camera \\ \hline
        Weight &  2.36kg\\ \hline
        Size &  60cm\\ \hline        
	\end{tabular}
	\label{table:Components}
\end{table}

The use of carbon fiber in the design of the frame imparts strength yet keeping it lightweight. Due to its low radar cross-section (RCS), carbon fiber is also challenging to detect by radar, making it the ideal material for such applications \cite{semkin2020analyzing}. The motor propeller combination can generate high thrust while utilizing minimal electrical power. Due to the UAV's high electrical efficiency, its endurance is increased. Pixhawk autopilot used here is the industry-leading platform for UAV control due to its stability, dependability, and precision, as well as its versatility and adaptability to various types of UAV operations. The Here 2 GPS receiver is able to provide precise position estimates, ensuring the accuracy and reliability of autonomous missions. Utilizing long range radio and telemetry modules enables the UAV to maintain long-range connectivity with the GCS and the operator. RFD900 is capable of transmission up to 40km, a high data transfer rate of up to 256 kbps, and low power consumption. It is designed with features such as error correction, automatic retransmission, and adaptive frequency hopping to ensure a stable and secure connection between the UAV and the ground station. The SJ4000 is a popular action camera with an ultra-wide 170$^\circ$ lens that can capture images and videos in high resolution. The camera is also waterproof, making it suitable for UAV use regardless of weather conditions.

\subsection{Range Estimation using Battery}
\label{Battery}
\begin{figure}[H]
	\centering
	\includegraphics[scale= 0.07]{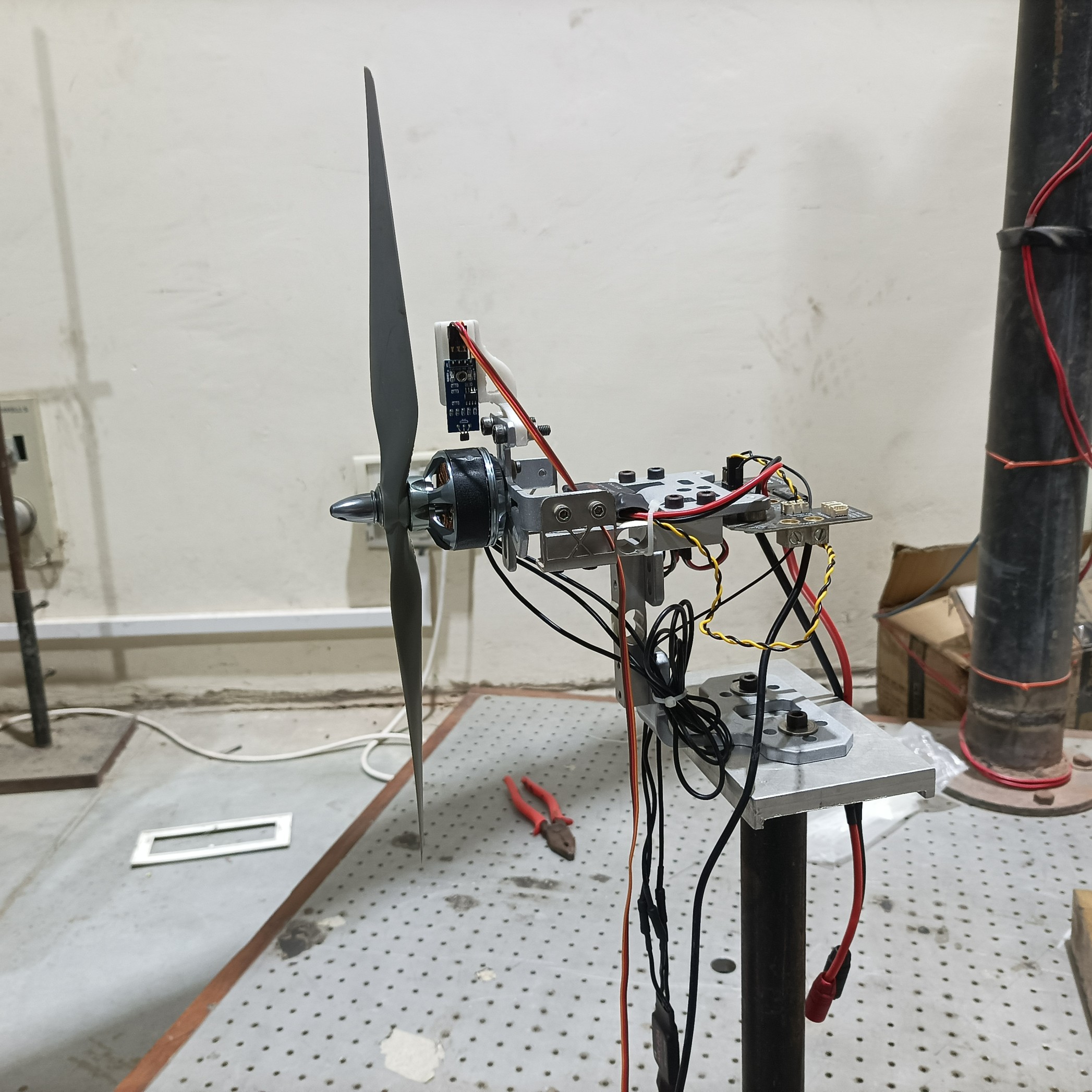}
	\caption{Propeller thrust test setup}
	\label{fig:Thrust}
\end{figure}

The recovered UAV is an electric-powered vehicle, and the first thing that must be performed is to determine its range. The approximate range assuming ideal conditions, can be determined based on the battery status. This helps to narrow down the search range for the possible source of origin. When the flight log data is corrupted or when data logging is manually disabled, this method is incredibly useful. The UAV was discovered carrying a four-cell LiPo battery which can be charged up to 16.8V. The total amount of energy contained in a completely charged battery is indicated by
\begin{equation}
\label{energy}
    E_{total} = V \frac{\text{Current rating (mAh)}}{1000} Wh
\end{equation}
The voltage reading of the battery when the UAV was retrieved is known from the physical examination to be 16.2V. A linear relation between battery voltage and discharge capacity (mAh) can be assumed in this region \cite{liu2022experimental} to get the discharge rating at 16.2V. Using Eq.~\eqref{energy}, the total energy used can be calculated as
\begin{equation*}
    E_{used} = E_{total}- E_{remaining}
\end{equation*}
\begin{equation*}
    E_{used} = \frac{16.8\times 6500 - 16.2\times 5300}{1000} = 23.34\; Wh
\end{equation*}
The power utilised by electronic components is necessary for estimating flight time and, by extension, range. The principal electrical power consumers onboard are the motors and telemetry module. The discovered camera is powered by a distinct battery, and no video transmitter was discovered. The power consumption of the telemetry module is detailed in the manufacturer-supplied datasheet. To calculate the electrical power consumed by the motor, we need to know the operating rpm of the motors so that they can generate sufficient propulsion for the vehicle's cruise speed. The vehicle's cruise speed in autonomous mode can be discovered in the autopilot's parameter list, which is detailed in subsequent sections. The thrust and electrical power required to attain cruise speed can be found on the datasheet for the motor. The official datasheets are frequently generated under ideal conditions; therefore, it is more reliable to determine the power consumption of the motor propeller combination through thrust testing as shown in Fig.~\ref{fig:Thrust}. To attain greater precision, flight log data can be analysed to determine the input PWM signals sent to the motors, and then thrust test data can be used to determine the power. The following equation can be used to compute the duration of the UAV's flight:
\begin{equation}
    T_{flight} =  \frac{E_{used}}{Total\;power\;consumed}\text{(hrs)}
\end{equation}
From the thrust test data, it was observed that each motor requires 94W of power to produce thrust for the UAV to attain an average forward speed of 2m/s, while the telemetry module consumes 5W. Using this, the vehicle's optimal flight time can be calculated as follows
\begin{equation*}
    T_{flight} =  \frac{23.34\times3600}{94\times4+5} = 220.5 sec
\end{equation*}
It should be noted that the calculated flight time and range presume ideal conditions. The actual values may be lower because wind disturbances, environmental conditions, operating temperature, noise, component deterioration over time, and other factors increase battery consumption.

The maximum range of the UAV can be given by 
\begin{equation}
    R_{max} =  T_{flight} V_{avg}
\end{equation}
\begin{equation*}
    R_{max} =  220.5\times2 = 441m
\end{equation*}


\subsection{Autopilot Flash Data Analysis}
\label{Flash}
\begin{figure}[ht]
	\centering
	\includegraphics[scale= 0.5]{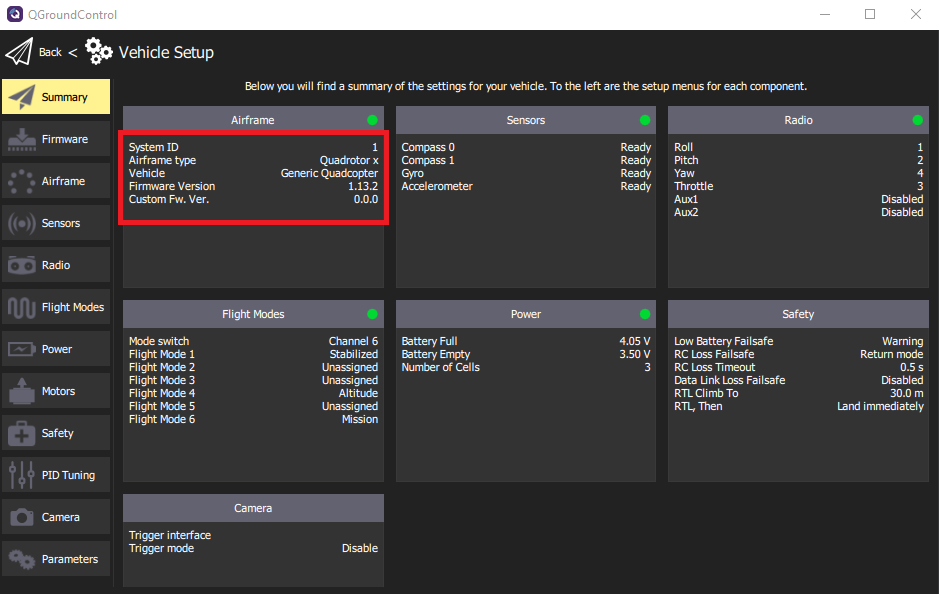}
	\caption{Firmware and airframe specifications of the UAV autopilot}
	\label{fig:Firmware}
\end{figure}

The Pixhawk Cube Orange has 2MB of flash memory for storing firmware, configuration parameters, bootloader, calibration data, and mission-specific information \cite{px4}. The Ground Control Station (GCS) software must be accessed in order to extract the UAV's firmware and other information. The GCS software provides an interface for the operator to monitor and control the UAV's flight and mission via wired or wireless communication. This makes it an essential component of any UAV system, allowing for the safe and efficient operation of the vehicle. QGroundControl (QGC) GCS software was utilized in this analysis \cite{qgc}. QGC offers complete flight management and mission planning for any MAVLink-enabled UAV. Using the correct COM port and baud rate, the autopilot can be accessed via a USB interface from the QGC. The firmware is intended to provide a software platform for tasks such as sensor data acquisition, motor control, advanced flight modes, external communication, and system-level function administration. Pixhawk operates with either the Ardupilot or Px4 firmware. The UAV's firmware specifications can be accessed from the vehicle setup section in QGC and are shown in Fig.~\ref{fig:Firmware}. The safety and failsafe measures employed by the UAV are enlisted in \autoref{table:Firmware}.

\begin{table}[H]
	\caption{Failsafe settings in the UAV} 
	\centering
	\begin{tabular}{|c|c|c|}
        \hline
        \textbf{Parameter} & \textbf{Value} & \textbf{Description}\\ \hline \xrowht{15pt}
        \multirow{2}{*}{Low battery failsafe trigger}&  Action: Warning & Warning when 
        the battery reaches the  \\ 
        &   Level: 15\%  & specified warn level\\ \hline
        Object detection & - & Disabled\\ \hline
        \multirow{3}{*}{RC loss failsafe trigger}&  Action: Return mode & The UAV will automatically return to the \\ 
        &   RC loss timeout: 2 sec & home point if the RC communication is \\
        & &  lost for more than 2sec\\ \hline
        Data link failsafe trigger & - & Telemetry failure does not activate a failsafe\\ \hline
        Geofence failsafe trigger& - & Disabled\\ \hline
        \multirow{2}{*}{Return to launch settings} & Altitude: 30m & UAV acquires an altitude of 30m, returns\\ 
        &  Post RTL: Land immediately  & to the home point and lands immediately\\\hline
        Land descent rate & 0.7m/s & -\\ \hline
        Vehicle telemetry logging& Enabled & Autopilot logs all the data on the storage device\\ \hline
	\end{tabular}
	\label{table:Firmware}
\end{table}

\begin{figure}[ht]
	\centering
	\includegraphics[scale= 0.42]{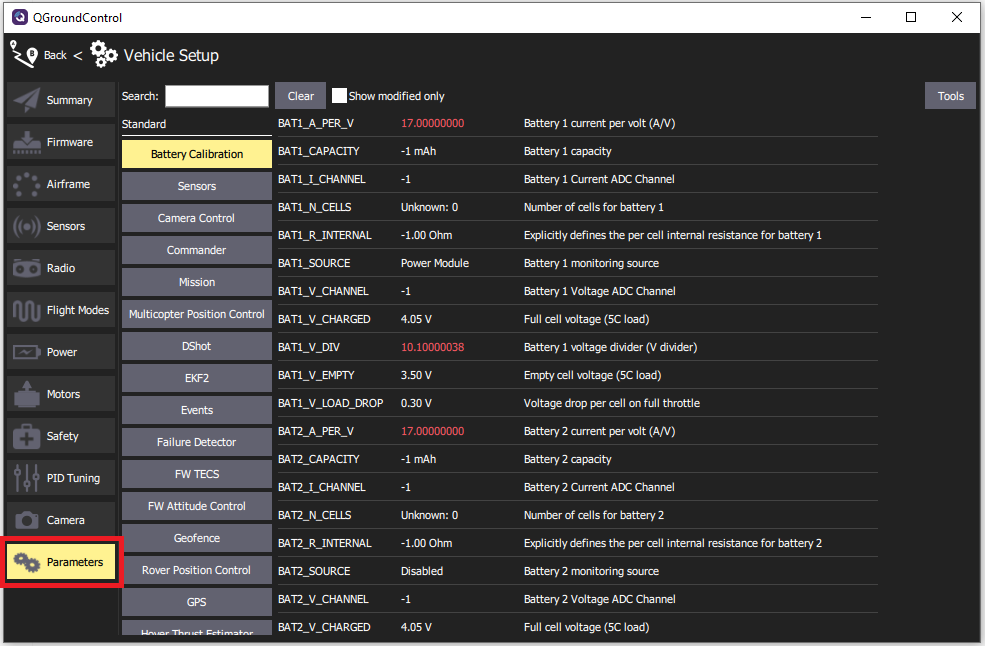}
	\caption{Autopilot parameter settings in the GCS software}
	\label{fig:Parameters}
\end{figure}

The configuration settings are parameters that specify the UAV's behavior in different situations. They can be modified to tailor the Pixhawk's behavior to a particular UAV. This includes the airframe type, flight mode parameters, sensor calibration values, etc. The configuration parameters can be accessed by browsing the full parameter list of the autopilot, as depicted in Fig.~\ref{fig:Parameters}. \autoref{table:Parameter} lists a number of crucial parameters for the UAV.

\begin{table}[H]
	\caption{Configuration parameters of the UAV} 
	\centering
	\begin{tabular}{|c|c|c|}
         \hline
		\textbf{Parameter} & \textbf{Value} & \textbf{Inference}\\ \hline
        COM\_ARM\_MIS\_REQ & 0 & UAV does not require valid mission to arm\\ \hline
        COM\_ARM\_WO\_GPS & 1 & Allows arming without GPS fix\\ \hline
        COM\_FLTMODE1& Stabilized & First flight mode\\ \hline
        COM\_FLTMODE4& Altitude & Fourth flight mode\\ \hline
        COM\_FLTMODE6& Mission & Sixth flight mode\\ \hline
        COM\_OBS\_AVOID & 0 & Obstacle avoidance disabled\\ \hline
        MIS\_LTRMIN\_ALT& 10m & Minimum loiter altitude\\ \hline
        MPC\_LAND\_SPEED& 0.7m/s & Landing descent rate \\ \hline
        MPC\_TILTMAX\_AIR& 45.0$^\circ$ & Maximum tilt angle in air \\ \hline
        MPC\_TKO\_SPEED& 1.50m/s & Takeoff climb rate\\ \hline
        MPC\_XY\_CRUISE& 5.0m/s &  Maximum horizontal velocity in mission\\ \hline
        MPC\_Z\_VEL\_MAX\_DN& 1.0m/s & Maximum vertical descent velocity\\ \hline
        MPC\_Z\_VEL\_MAX\_UP& 3.0m/s & Maximum vertical descent velocity\\ \hline
        \multirow{2}{*}{FD\_ESCS\_EN} & \multirow{2}{*}{1} & Enable failure detector check on ESCs that\\
        &  & report the arming state\\ \hline
        \multirow{2}{*}{GPS\_1\_GNSS} & \multirow{2}{*}{0} & GNSS systems for the primary receiver.\\
        &  & Use GPS system (with QZSS)\\ \hline
        GPS\_1\_PROTOCOL& u-blox & Protocol for primary GPS\\ \hline
        MAV\_TYPE& Quadrotor & MAVLink airframe type\\ \hline
        PWM\_AUX\_RATE& 50Hz & PWM output frequency of the auxiliary channels\\ \hline
        \multirow{2}{*}{SDLOG\_BOOT\_BAT} & \multirow{2}{*}{0} & Logging will start from boot even if\\
        &  & battery power is not detected\\ \hline
        SDLOG\_MODE& When armed until disarmed & Logging mode\\ \hline
        SYS\_AUTOCONFIG& Keep parameters & Automatically configure the values\\ \hline
        SYS\_MC\_EST\_GROUP& ekf2 & Set multicopter estimator group\\ \hline

	\end{tabular}
	\label{table:Parameter}
\end{table}

\begin{figure}[H]
	\centering
	\includegraphics[scale= 0.45]{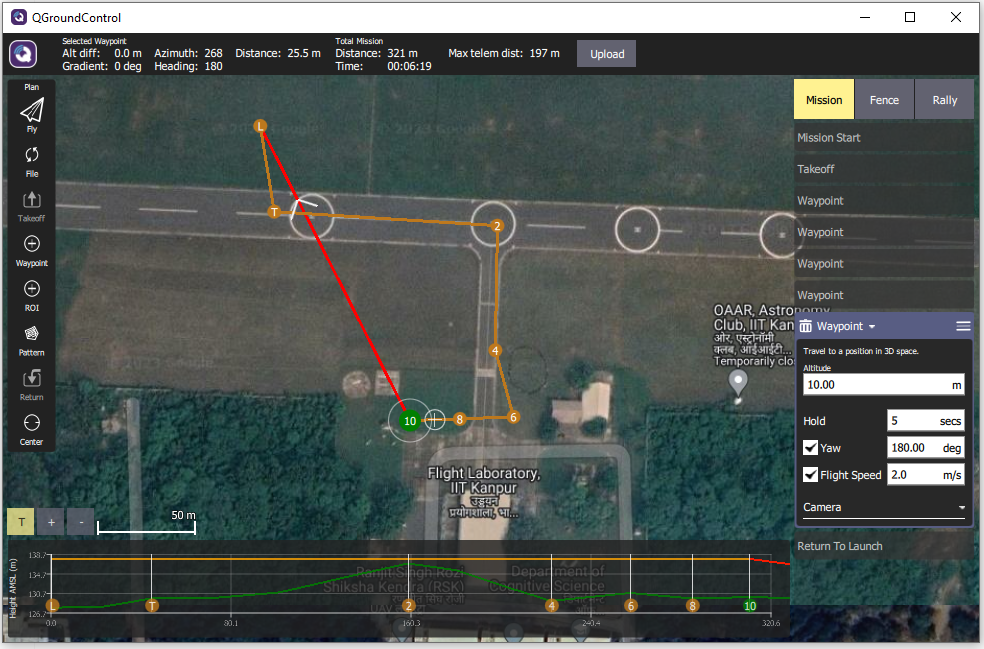}
	\caption{UAV's pre-planned mission retrieved from the autopilot flash memory}
	\label{fig:Mission}
\end{figure}

The mission specifics can be used to retrieve autonomous mission details such as the home location, takeoff and landing points, waypoints, flight path, and failsafes. This aids in determining the possible origin, flight path, and other objectives of the UAV. Fig.~\ref{fig:Mission} depicts how the pre-planned mission details can be read directly from flash memory via the flight plan section of the QGC software. The location marked `L' represents the home or launch point of the UAV, while the location marked `T' represents the launching point. The numbers and the line connecting them depict, respectively, the location of the waypoints and the path the UAV is instructed to follow. It should be noted that the number of waypoints does not increase at regular intervals. This is due to the fact that the UAV is commanded to perform additional tasks at these locations, such as velocity changes, loitering, etc.

The UAV is initially instructed to take off at the height of 10m and move along the runway to waypoint 2 at a speed of 5m/s (specified in \autoref{table:Parameter}). It then enters the taxiway and pursues the remaining path at a speed of 2m/s. The waypoint description indicates that between waypoints 6 and 10, the UAV must sustain a heading of 180 degrees and hold the position at the tarmac's edge (visible in Fig.~\ref{fig:Mission}) for 5sec. The holding period is established so that a sufficient quantity of high-quality images can be collected at these locations. Once the UAV has collected data from all the specified waypoints, it must return to the launch location by retracing the path marked by the red line.

\subsection{Flight Log Analysis}
\begin{figure}[H]
	\centering
	\includegraphics[scale= 0.5]{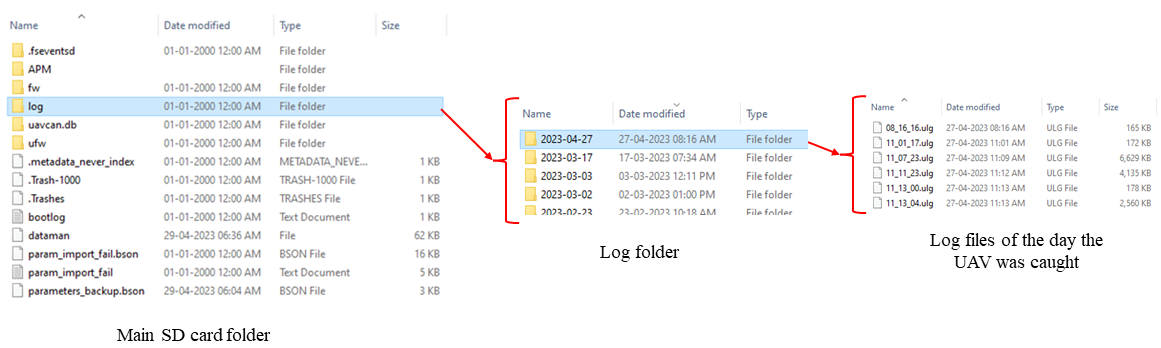}
	\caption{Log file retrieval procedure}
	\label{fig:Log}
\end{figure}

Pixhawk has a logging feature that generates log files used for debugging and analyzing the UAV's flight behavior. Log files include information such as sensor readings, vehicle status, and control inputs that were recorded during the flight. Using the GCS software, the binary data can be converted to a readable format. Due to their large size, log files cannot be saved in the flash memory. Pixhawk employs a microSD card to store flight log data in addition to the boot log file, a backup copy of configuration parameters, and other crucial information. To access the log files, the SD card must be removed from the Pixhawk and accessed. Fig.\ref{fig:Log} depicts the steps required to access a particular log file.

The log files are in ULG format and can be analyzed using Matlab Flight Log Analyzer, Px4 Flight Review, or any other third-party application. Matlab Flight Log Analyzer was utilized for this analysis due to its advantages in terms of customization, data synchronization, interactive data exploration, and data exporting. \autoref{table:Flight Summary} summarizes the UAV's entire flight course. 

\begin{table}[H]
	\caption{Flight summary} 
	\centering
	\begin{tabular}{|c|c|}
         \hline
        Total flight time & 2min 26sec\\ \hline
        Total distance & 295.1m\\ \hline
        OS Version & NuttX, v11.0.0\\ \hline
        Estimator & EKF2\\ \hline
        Average speed & 2m/s\\ \hline
        Max speed & 5.06m/s\\ \hline
        Max up speed & 2.8m/s\\ \hline
        Max down speed & 1.25m/s\\ \hline
        Max tilt angle & 19.8$^\circ$\\ \hline    
	\end{tabular}
	\label{table:Flight Summary}
\end{table} 

\begin{figure}[H]
\centering
\begin{subfigure}{.54\textwidth}
  \centering
  \includegraphics[scale= 0.32,]{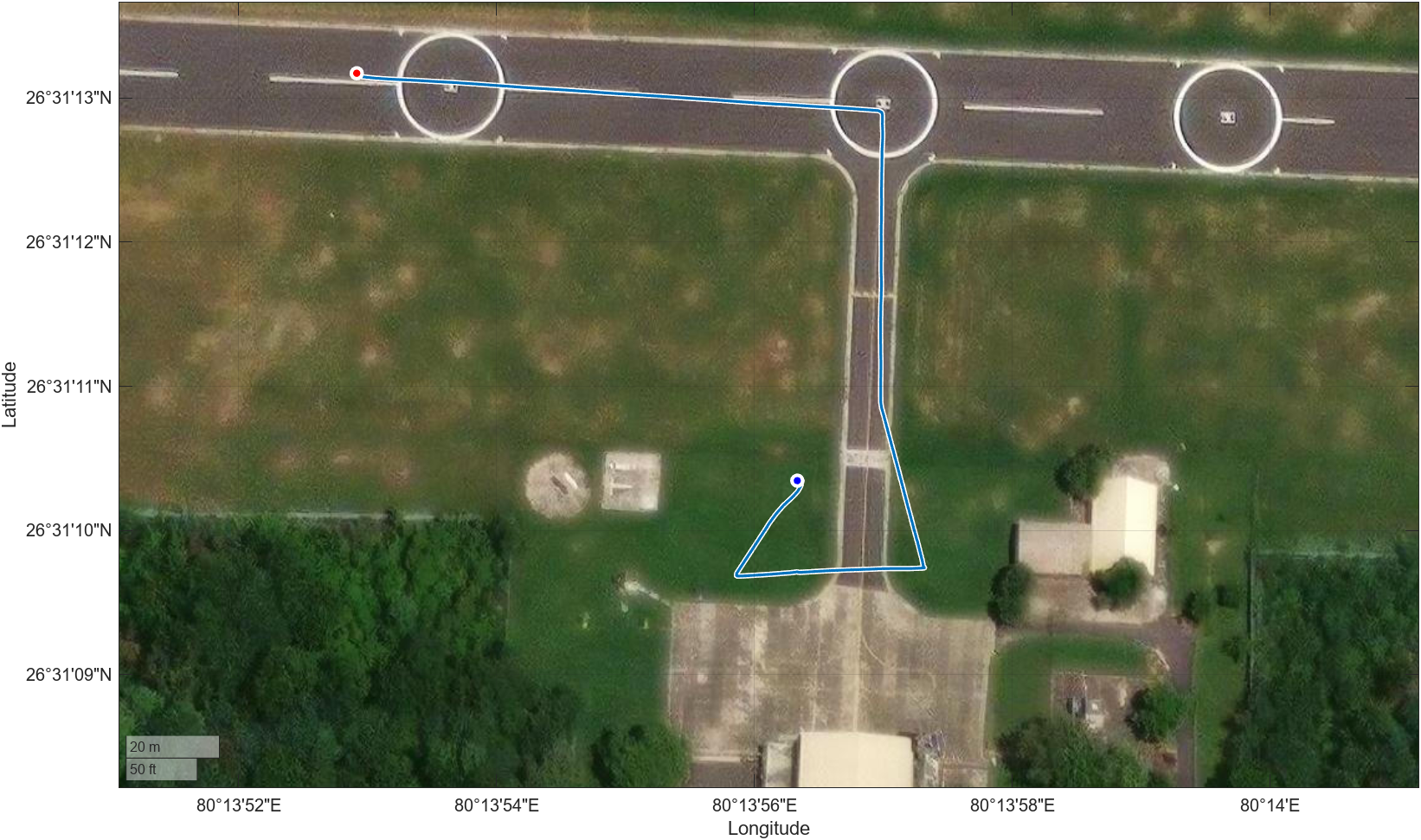}
  \caption{Map view}
  \label{fig:FltPath1}
\end{subfigure}
\begin{subfigure}{.45\textwidth}
  \centering
  \includegraphics[scale= 0.475,]{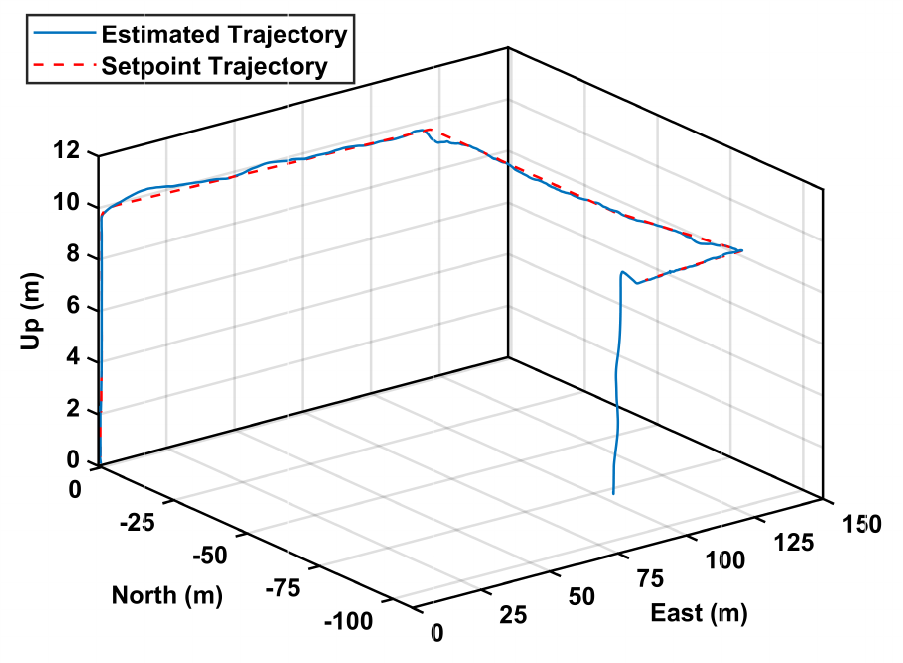}
  \caption{3D trajectory with takeoff point as the origin}
  \label{fig:FltPath12}
\end{subfigure}
\caption{Flight path of the UAV}
\label{fig:Flight path}
\end{figure}

The trajectory followed by the UAV is shown in Fig.~\ref{fig:Flight path}. The autopilot integrates the barometer altitude and GPS altitude to provide a more precise fused altitude estimation. Similarly, precise x-y coordinates are estimated by combining GPS and IMU data. As discussed in \autoref{Flash}, the UAV does not successfully complete its mission by returning to the launch position. This is because the UAV was intercepted and captured as it approached waypoint 10 in Fig.~\ref{fig:Mission}. 

\begin{figure}[ht]
\centering
\begin{minipage}{.49\textwidth}
  \centering
  \includegraphics[scale= 0.5]{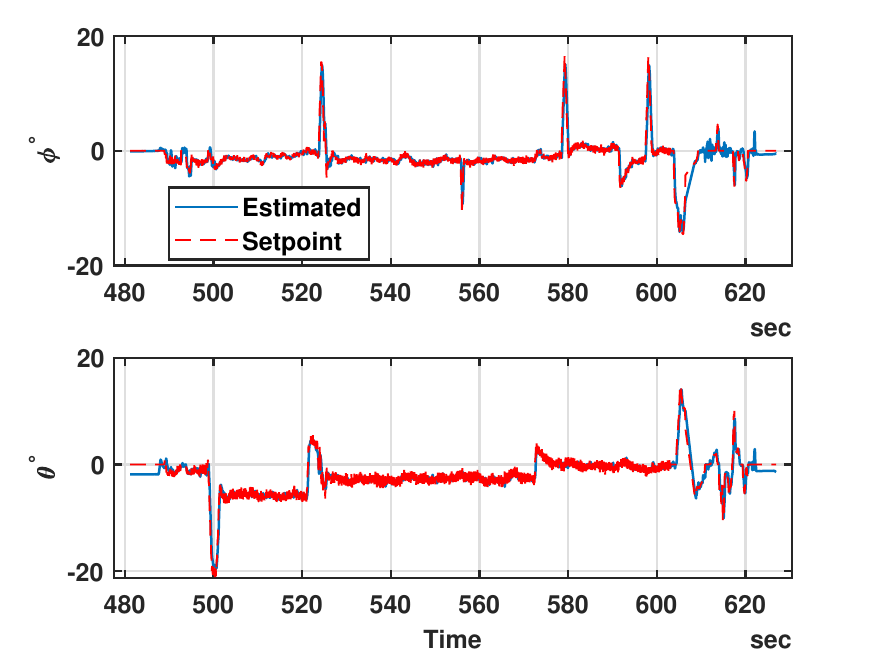}
  \caption{Vehicle attitude}
  \label{fig:Attitude}
\end{minipage}%
\begin{minipage}{.49\textwidth}
  \centering
  \includegraphics[scale= 0.5]{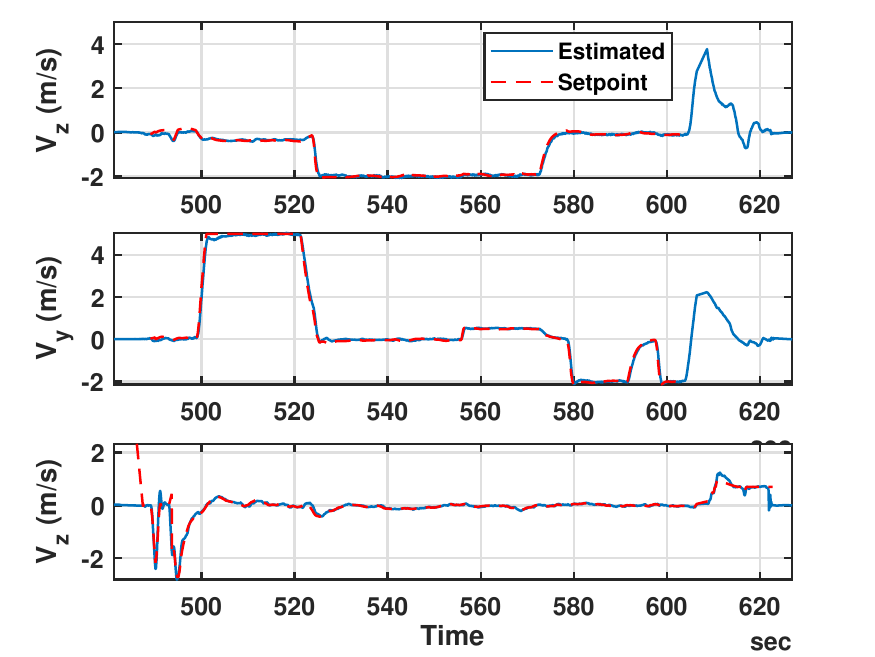}
  \caption{Vehicle velocity in the body frame}
  \label{fig:Velocity}
\end{minipage}
\end{figure}

 Fig.~\ref{fig:Attitude} and Fig.~\ref{fig:Velocity} depict the attitude and x-y-z velocity of the UAV in the body frame, respectively. Due to the low magnitude of the reference velocities, the UAV's attitude remains small for the majority of flight time. The erratic peaks observed in the attitude are the result of unanticipated wind turbulence. Notably, the time axis in these graphs does not begin at 0 seconds. This is because the Pixhawk autopilot's clock begins when the system is powered on, whereas data recording does not begin until the UAV is armed. Therefore, it can be inferred that the UAV was armed and initiated the mission after approximately 481 seconds of being powered on.

Since the problem statement necessitates tracing the precise flight path of the surveillance UAV, only the pertinent flight log data and corresponding results have been discussed. In addition to the data and results discussed earlier, the log files also contain information such as attitude rate, angular velocity and acceleration, radio inputs, actuator outputs, vibration metrics, GPS uncertainty, magnetic field, logged messages, etc. The flight history of the UAV can be more accurately estimated by combining these data.

\subsection{Other Data Retrieved}
\begin{figure}[H]
\centering
\begin{subfigure}{.49\textwidth}
  \centering
  \includegraphics[scale= 0.1,]{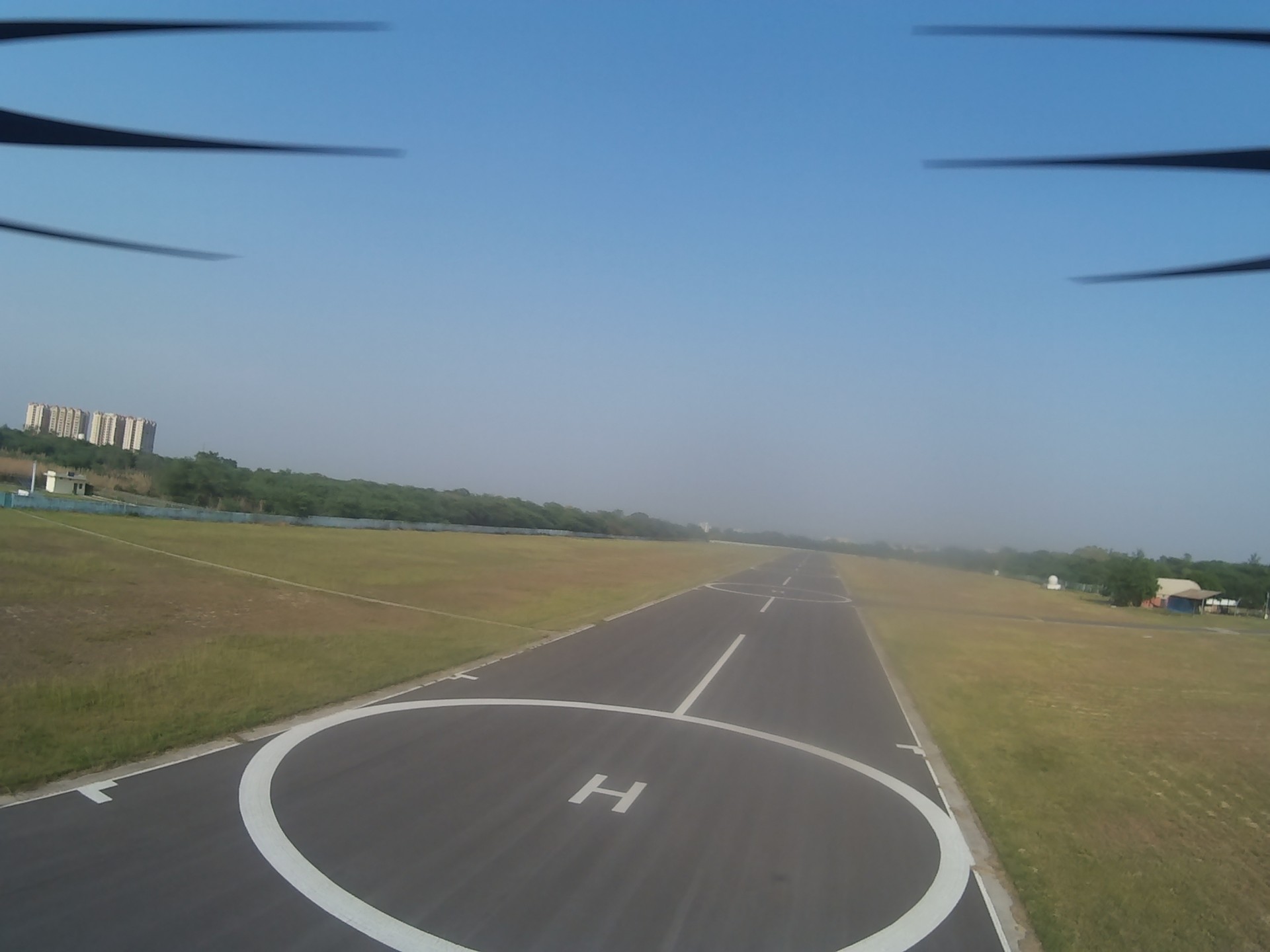}
  \caption{Image 1}
  \label{fig:Flight1}
\end{subfigure}
\begin{subfigure}{.49\textwidth}
  \centering
  \includegraphics[scale= 0.1,]{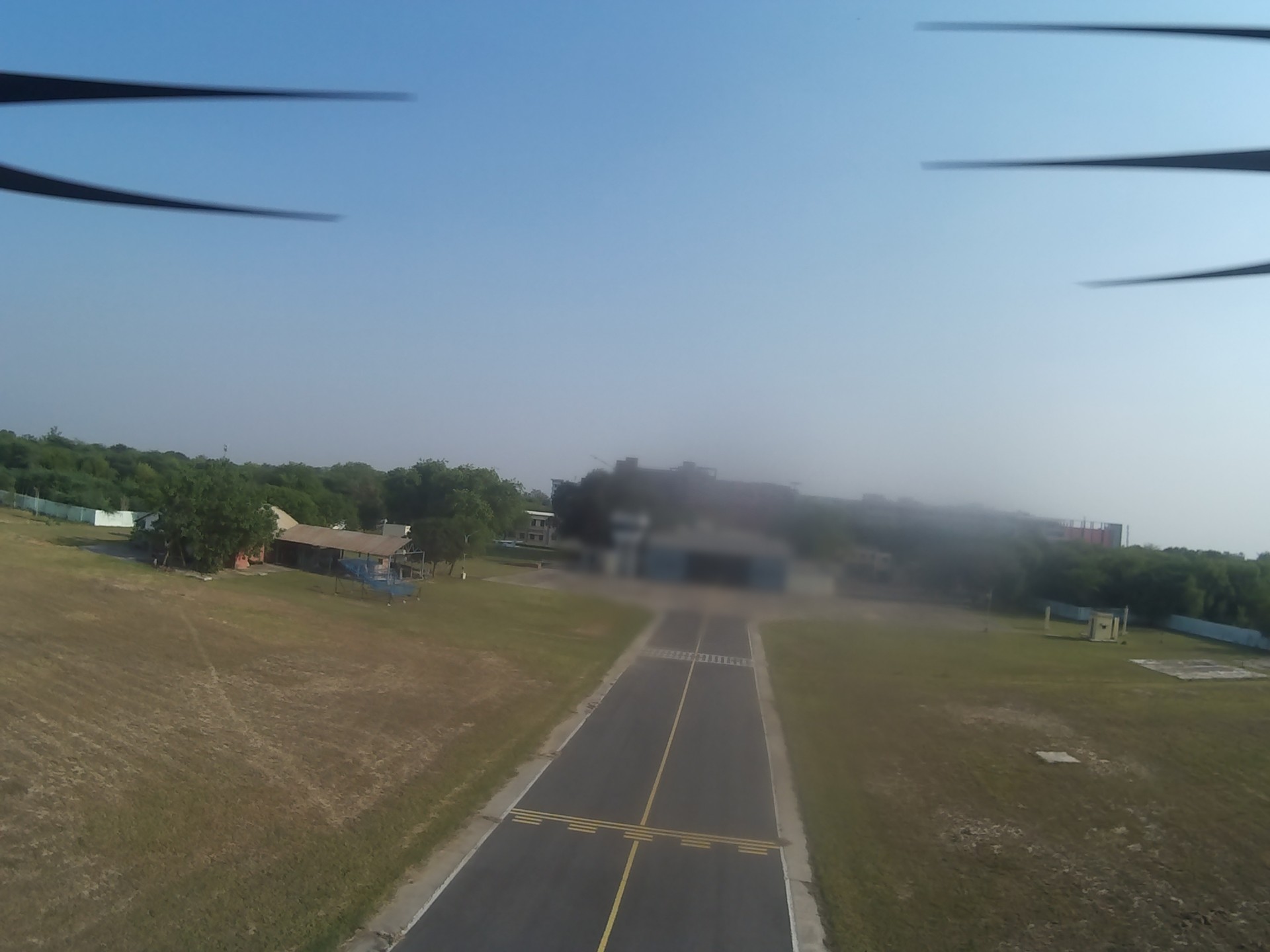}
  \caption{Image 2}
  \label{fig:Flight2}
\end{subfigure}

\begin{subfigure}{.49\textwidth}
  \centering
  \includegraphics[scale= 0.1]{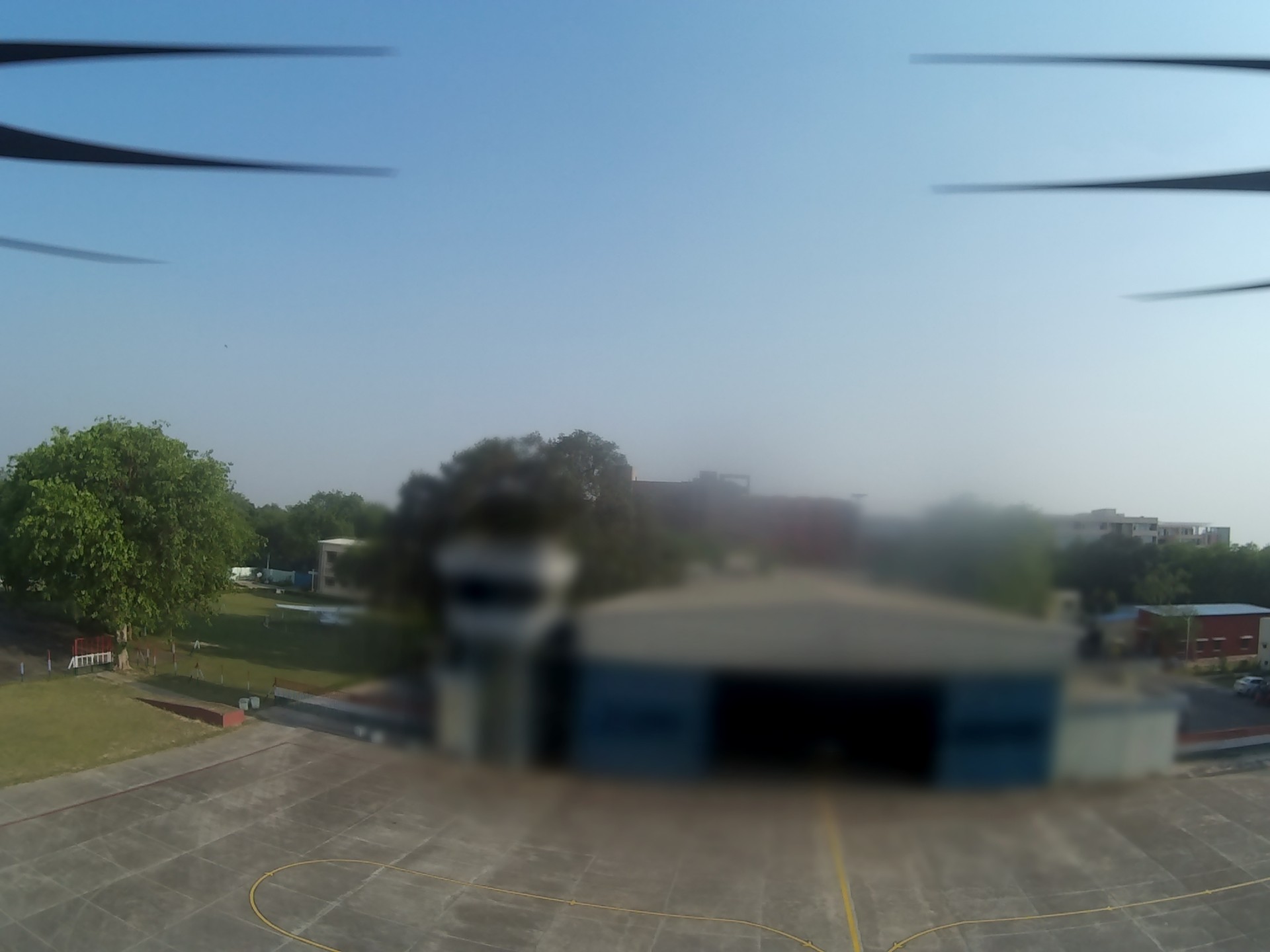}
  \caption{Image 3}
  \label{fig:Flight3}
\end{subfigure}
\begin{subfigure}{.49\textwidth}
  \centering
  \includegraphics[scale= 0.1,]{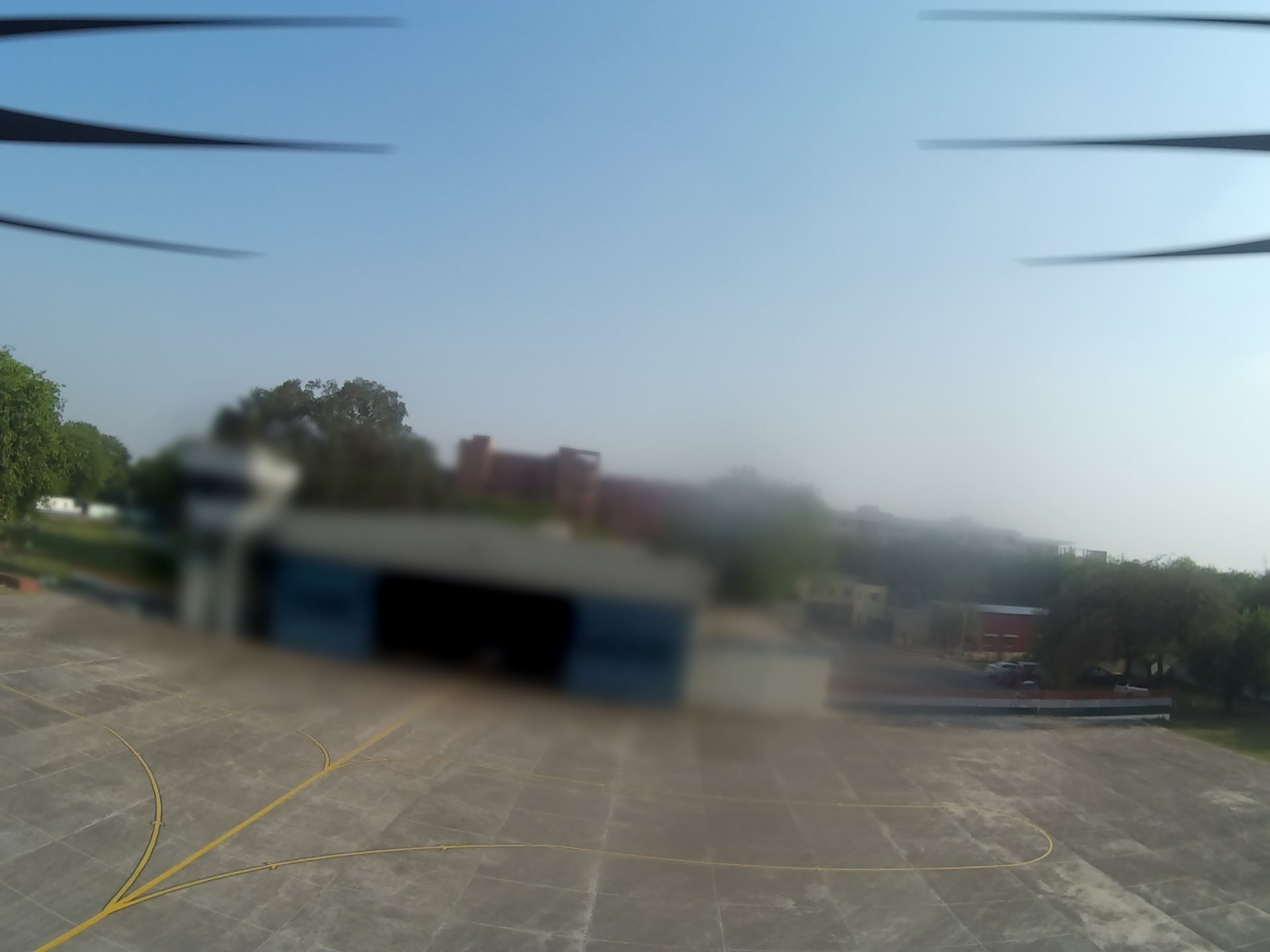}
  \caption{Image 4}
  \label{fig:Flight4}
\end{subfigure}
\caption[]{Images extracted from the camera onboard the UAV \footnotemark[5]}
\label{fig:Camera}
\end{figure}

After a comprehensive analysis of the autopilot data, it is imperative to search for any additional storage devices on the UAV, as the data may provide additional information about the UAV's whereabouts, origin, and operator. As specified in section \ref{Physical Examination}, the UAV is equipped with an HD camera for gathering images. The camera includes a 16GB micro SD card because high-resolution images and videos necessitate a secondary large storage medium. A thorough examination of the SD card revealed that the UAV captured numerous images from a variety of locations and perspectives during its flight, some of which may contain sensitive information. Several of these images are depicted in Fig.~\ref{fig:Camera}. Although the images are not geo-tagged and identifying the individuals captured by the camera is difficult, the time stamp of the images can be combined with the flight log data to generate a visual representation of the UAV's flight path. This can aid in accurately pinpointing the origin of the UAV.

\footnotetext[5]{Images blurred due to security concerns.}

\section{Conclusion}
\label{Conclusion}
This study highlights the significance of UAV forensics as well as the associated difficulties. As the use of UAVs continues to expand in multiple industries, there is an increasing need for forensic investigation capabilities to ensure the safe and responsible operation of these systems. Using a custom reconnaissance UAV as a case study, a detailed methodology for the forensic analysis of a UAV has been outlined.  The novel contribution of this study is the use of a cheaper custom UAV for the analysis, as they can be tailored to a particular application, making them difficult to monitor and neutralize. This was accomplished by analyzing the various components of the recovered UAV in chronological order using open-source software and MATLAB. The analysis contributes novel insights to the existing corpus of knowledge on drone forensics. Through a combination of physical analysis, data recovery, and data analysis, we have demonstrated that UAV forensics has the potential to yield invaluable insights into the operation of these complex systems. The data has been meticulously analyzed, and conclusions have been derived based on evidence from a variety of sources. The results indicate that forensic analysis of UAVs can aid in the investigation of accidents, incidents, and criminal activities involving these systems. However, UAV forensics is a difficult and rapidly evolving field that requires additional research and development to enhance the efficacy and dependability of forensic techniques.

\bibliography{sample}

\end{document}